\documentclass[aps,prstab,twocolumn,superscriptaddress]{revtex4-1}

\usepackage{graphicx}
\usepackage{float}
\usepackage{amsmath}
\usepackage{gensymb}
\usepackage[caption=false]{subfig}
\begin{document}


\title{Beam-induced Back-streaming Electron Suppression Analysis for Accelerator Type Neutron Generators}

\author{Cory Waltz}
\affiliation{Department of Nuclear Engineering, University of California, Berkeley, CA 94720, USA}
\affiliation{Lawrence Livermore National Laboratory, Livermore, CA 94550, USA}
\author{Mauricio Ayllon}
\affiliation{Department of Nuclear Engineering, University of California, Berkeley,  CA 94720, USA}
\author{Tim Becker}
\affiliation{Berkeley Geochronology Center, 2455 Ridge Road,Berkeley, CA 94720, USA}
\author{Lee Bernstein}
\author{Ka-Ngo Leung}
\author{Leo Kirsch}
\affiliation{Department of Nuclear Engineering, University of California, Berkeley, CA 94720, USA}
\author{Paul Renne}
\affiliation{Berkeley Geochronology Center, 2455 Ridge Road,Berkeley, CA 94720, USA}
\author{Karl Van Bibber}
\affiliation{Department of Nuclear Engineering, University of California, Berkeley, CA 94720, USA}

\date{\today}

\begin{abstract}
A facility based on a next-generation, high-flux D-D neutron generator has been commissioned and it is now operational at the University of California, Berkeley. The current generator design produces near monoenergetic 2.45 MeV neutrons at outputs of 10$^{8}$ n/s. Calculations provided show that future conditioning at higher currents and voltages will allow for a production rate over 10$^{10}$ n/s. A significant problem encountered was beam-induced electron backstreaming, that needed to be resolved to achieve meaningful beam currents. Two methods of suppressing secondary electrons resulting from the deuterium beam striking the target were tested: the application of static electric and magnetic fields. Computational simulations of both techniques were done using a finite element analysis in COMSOL Multiphysics\textsuperscript{\textregistered}. Experimental tests verified these simulation results. The most reliable suppression was achieved via the implementation of an electrostatic shroud with a voltage offset of -800 V relative to the target.   
\end{abstract}

\pacs{}

\maketitle

\section{Introduction}
\begin{figure}
	\centering
	\includegraphics[width=0.50\textwidth]{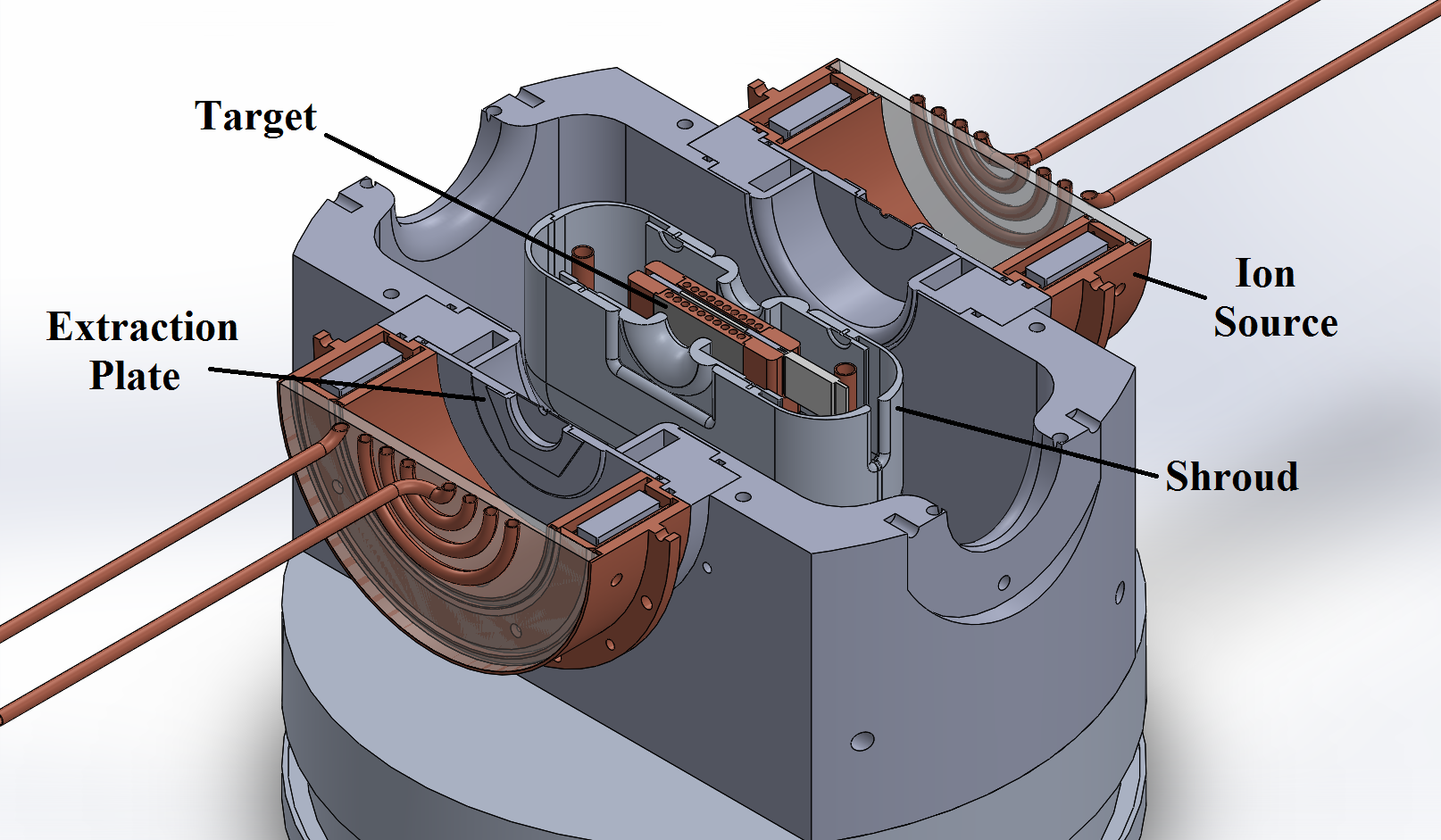}
	\caption{HFNG cross section exposing target, shroud, and ion sources. For scale, the ion source is approximately 19 cm in diameter.}
	\label{fig:hfng_cs}
\end{figure}

The High Flux Neutron Generator (HFNG) located at the University of California Berkeley is designed around two radio frequency-driven multi-cusp ion sources that straddle a titanium-coated water cooled copper target, as shown in Figure \ref{fig:hfng_cs}. Positively charged deuteron ions are accelerated up to 125 keV from the ion sources and self-load into the target. Upon target saturation, neutron generation occurs through the d(d,n)\textsuperscript{3}He fusion reaction. In order to take advantage of the fact that the highest neutron flux is in the forward direction with respect to the beam, a sample holder slot is located in the center of the target at a distance of 8 mm from the location where the deuterium ions strike the target and generate neutrons. A detailed description of the design and operation of the HFNG is being prepared for publication.

As deuterium ions extracted from the ion source strike the target, ionization occurs at the surface, releasing secondary electrons. According to a study done in \cite{Electron_emission}, approximately 1.2 electrons are emitted per deuterium ion striking a titanium target at 100 keV. Secondary electrons emitted from metal surfaces have energies around 10 eV, and typically not more than 30 eV \cite{SE_energy}. Secondary electrons accelerate away from the target due to the negative electric potential, resulting in an electron beam that strikes the extraction plate. The collision of this secondary electron beam with the extraction plate result in the emission of bremsstrahlung X-rays. If the current density is large enough, melting can occur. Collision of electrons with any insulators inside the vacuum chamber will result in charge build-up, which over time will discharge in the form of an arc. If arcing is frequent, damage to the high voltage power supply can occur. Lastly, secondary electrons represent a leakage current to the high voltage power supply, making it difficult to accurately determine the contribution of current from deuterons compared to secondary electrons. Suppression of these electrons is vital for continuous operation of a neutron generator. Two methods of suppressing electrons were tested on the HFNG: using magnetic fields produced by permanent magnets and using an electric field created by installing an electrostatic shroud. Analysis results were first published in \cite{waltzPHD}.

\section{COMSOL Multiphysics\textsuperscript{\textregistered} Finite Element Simulation Overview}

COMSOL Multiphysics\textsuperscript{\textregistered} was used to model the electrical and magnetic fields within the HFNG vacuum chamber \cite{Comsol}. COMSOL Multiphysics\textsuperscript{\textregistered} uses the macroscopic form of Maxwell's equations to determine the magnetic and electric fields.

\begin{figure}
	\centering
	\hspace*{\fill}
	\subfloat[\label{subfig:target_mag}]{\includegraphics[width=0.35\textwidth]{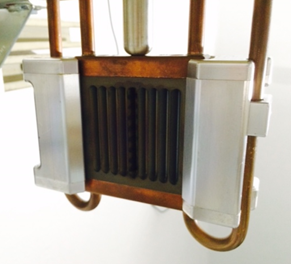}}
	\hfill
	\subfloat[\label{subfig:mag_cad}]{\includegraphics[width=0.11\textwidth]{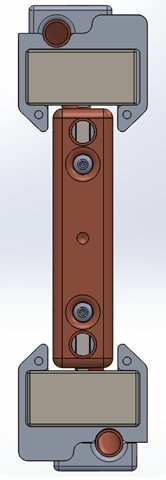}}
	\hspace*{\fill}
	\caption{Neodymium magnet holders. (a) Photo of the magnet holders clamped to target cooling tubes. (b) CAD drawing, top view, showing neodymium magnets.}
	\label{fig:target_magnets}
\end{figure}

\noindent The fields are computed by using a finite element analysis over a meshed geometry. User entered boundary conditions include the electric potentials at the surface of a material, and the remanent flux density (B$_{r}$) of magnets. Charged particle trajectories can then be computed by COMSOL Multiphysics\textsuperscript{\textregistered} by applying the forces on the particles due to the calculated magnetic and electric fields. Corrections to the electric field are made using an iterative process to fold in the effect of space charge of the beam. The simulations were performed under non-relativistic conditions since relativistic effects are negligible within this energy range.

\section{Magnetic Suppression}

One method of electron suppression involves the use of magnetic fields.  Figure \ref{fig:target_magnets} shows the implementation of large neodymium magnets on the HFNG target. The magnets are aligned with opposite poles facing each other, creating magnetic field lines parallel with the target surface. This causes ejected electrons to spiral around the field lines and back toward the target surface. It is worth noticing that the electron does not only feel the force due to the magnetic field, but it also feels the force of the electric field used to accelerate the deuteron ions. This net force results in an E$\times$B drift, which causes an orthogonal drift of the electron with respect to the magnetic and electric fields. 

\begin{figure}
	\centering
	\subfloat[\label{subfig:mag_field}]{\includegraphics[width=0.4\textwidth]{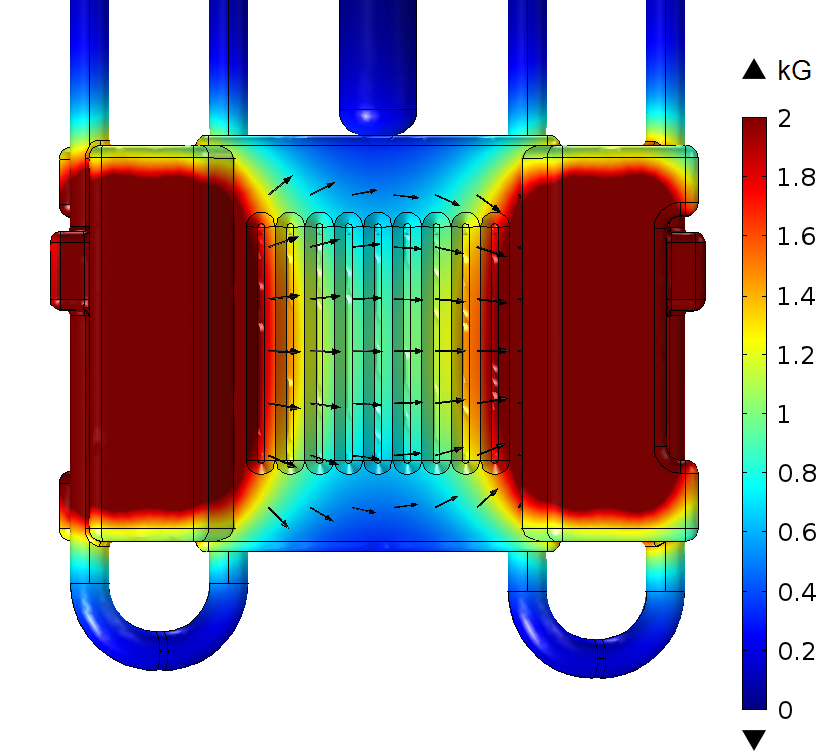}}
	\hfill
	\subfloat[\label{subfig:elec_field}]{\includegraphics[width=0.4\textwidth]{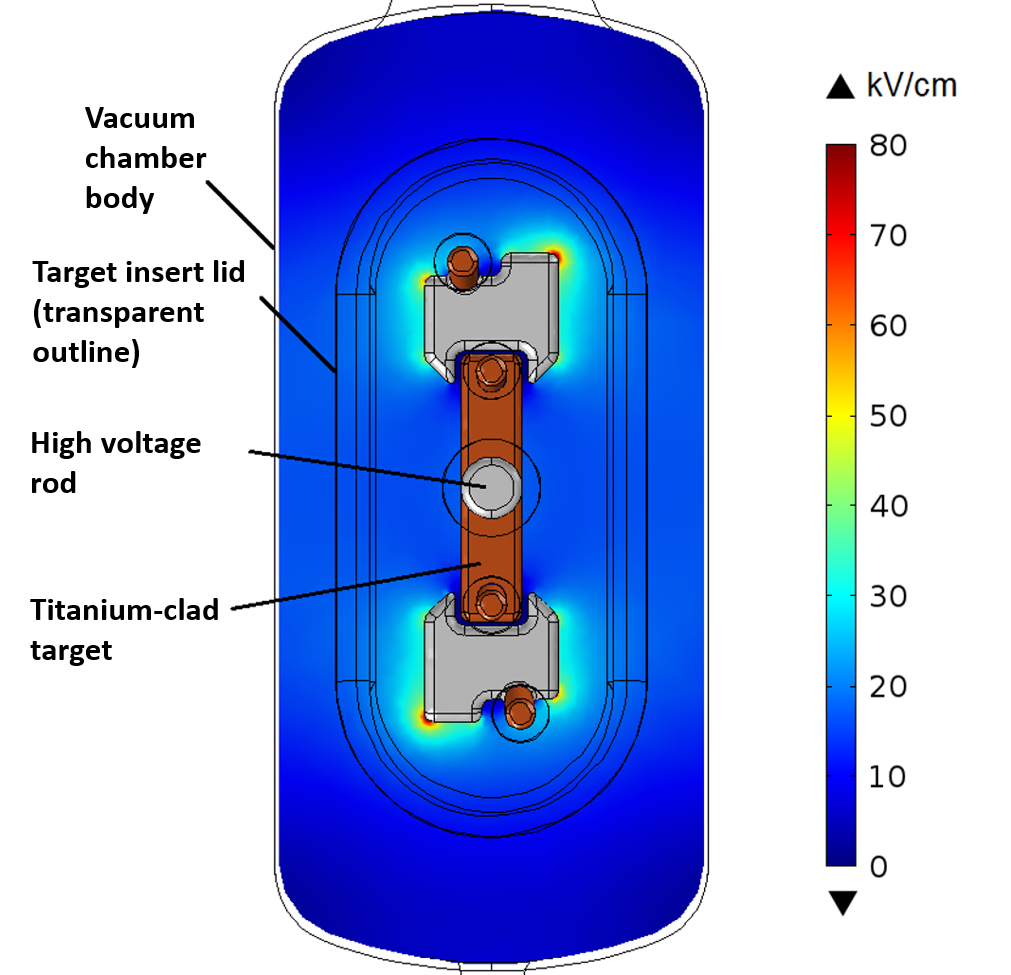}}
	\caption{Simulations of the (a) magnetic fields and (b) electric field}
	\label{fig:mag_and_elec_fields}
\end{figure}

\subsection{Electric and Magnetic Field Simulation Results}

A COMSOL Multiphysics\textsuperscript{\textregistered} simulation shows the strength of the magnetic field across the surface of the target in Figure \ref{subfig:mag_field}. The neodymium magnets were grade N50, giving them a $B_{r}$ between 14.1-14.5 kG. The black arrows represent the direction of the magnetic field lines in the respective arrow location. The magnetic field strength at the center of the target is approximately 800 gauss. The field strength along the surface was verified using a Hall probe. The strength of the electrical field can be seen in Figure \ref{subfig:elec_field}, which shows a top view of the target assembly. The field is quite large near the corners of the magnet holders, approaching a value of 80 kV/cm. It is important to remain below the breakdown voltage in vacuum, which can be approximated by the following empirical relation \cite{HV_Breakdown}:

\begin{equation}
E_{max}\left[\frac{kV}{cm}\right] = \frac{8000}{V\left[kV\right]}
\end{equation}

where $V$ is the extraction voltage and $E_{max}$ is the maximum magnitude of electric field allowed. At 100 kV, the breakdown E-field becomes approximately 80 kV/cm.

\subsection{Simulation of Electron Trajectories Using Magnetic Suppression}

\begin{figure}
	\centering
	\includegraphics[width=0.5\textwidth]{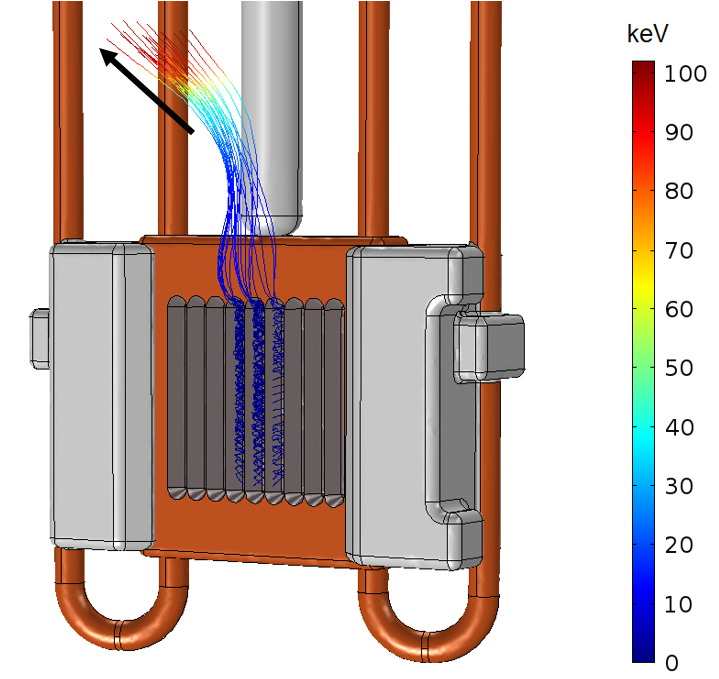}
	\caption{COMSOL Multiphysics\textsuperscript{\textregistered} electron path simulation}
	\label{fig:comsol_electrons}
\end{figure}

Figure \ref{fig:comsol_electrons} shows the COMSOL Multiphysics\textsuperscript{\textregistered} simulated electron paths, assuming three stripes of deuteron beams striking the center of the angled target.

The voltage of the target is -100 kV. The electrons liberated by the deuteron beams spiral up the target due to the E$\times$B drift.  When the electrons reach an area of weakened magnetic field, the electric field dominates and the electrons accelerate to the vacuum chamber wall. Experimental tests confirmed the electrons were colliding with the vacuum chamber wall exactly in the location shown by the simulation. 

Further analyzing the simulation shows that when the electrons spiral back toward the target, they get close to the surface but do not collide. The magnetic force a particle feels is proportional to its velocity. The electron slows down as it spirals back toward the surface due to the electric field, causing the magnetic force to weaken. Ultimately, this is a consequence of the fact that as magnetic fields can only change the path of the electron and do not add energy, it is difficult for the electrons to overcome the electric field force and recombine with the target.  

\section{Electrostatic Shroud}
An electrostatic shroud  is a device that suppresses back-streaming electrons by introducing an opposing electric field near the target surface. This is done by installing a shroud around the target that is kept at a more negative potential. Locally, the resulting electric field is directed away from the target surface, causing electrons to be repelled from the shroud and return to the target.

\begin{figure}
	\centering
	\includegraphics[width=0.5\textwidth]{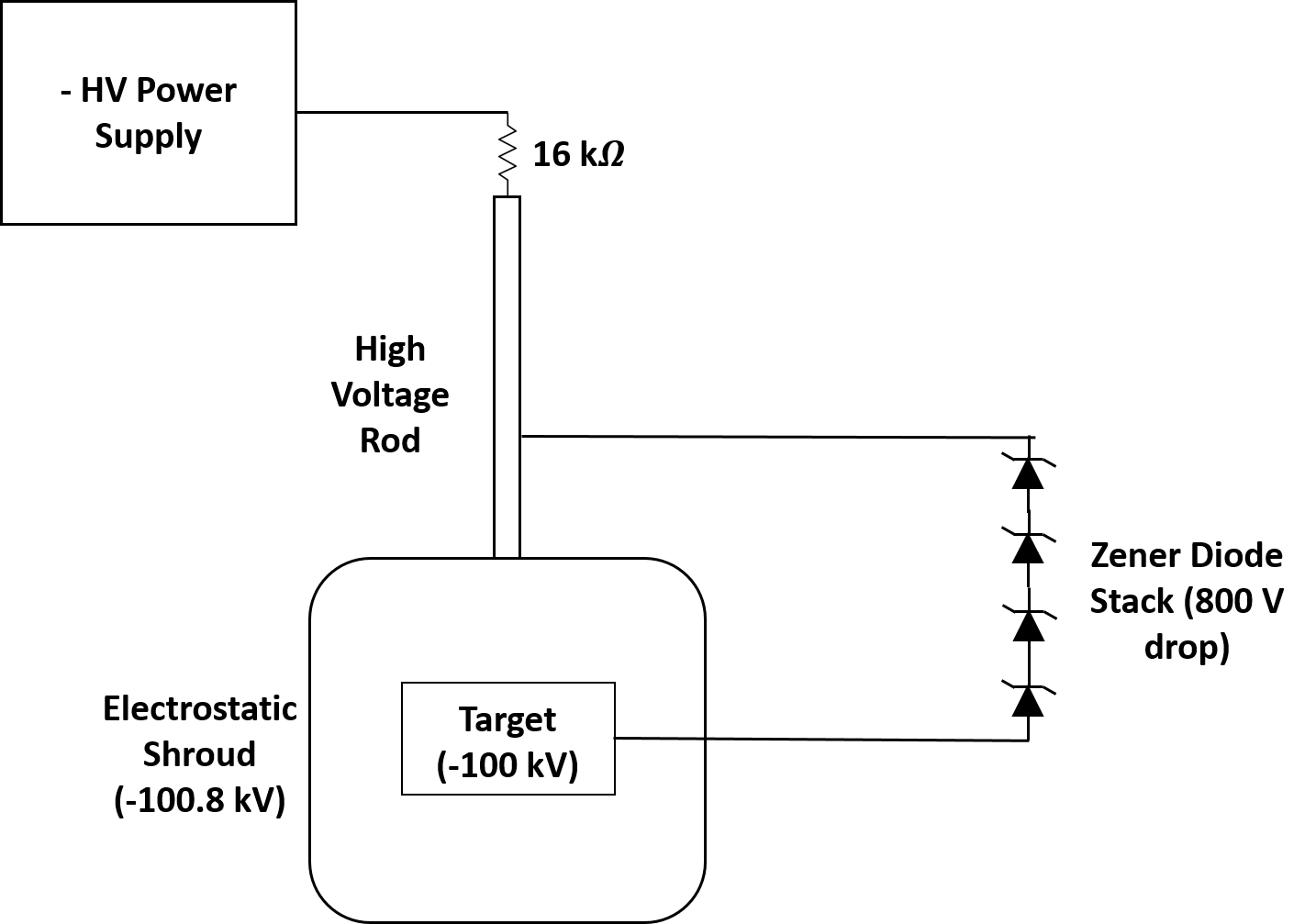}
	\caption{Circuit schematic of the diode assembly showing the 16 $k\ohm$ resistor that reduces current flow during arcing in order to protect the machine}
	\label{fig:diodes}
\end{figure}

The shroud design used for the HFNG is shown in Figure \ref{fig:shroud}. The clamshell design allows for easy removal.  Exchangeable face-plates allow for the ability to easily change the beam entrance window geometry. The voltage difference between the shroud and target is created through the use of multiple zener diodes in series, as shown in Figure~\ref{fig:diodes}. The total breakdown voltage of the zener diode chain was made as high as 2400 V.

\subsection{Electric Field Simulation Results}

\begin{figure}
	\subfloat[\label{subfig:closed_shroud}]{\includegraphics[width=0.14\textwidth]{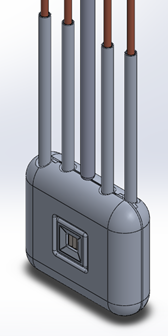}}
	\hfill
	\subfloat[\label{subfig:open_shroud}]{\includegraphics[width=0.33\textwidth]{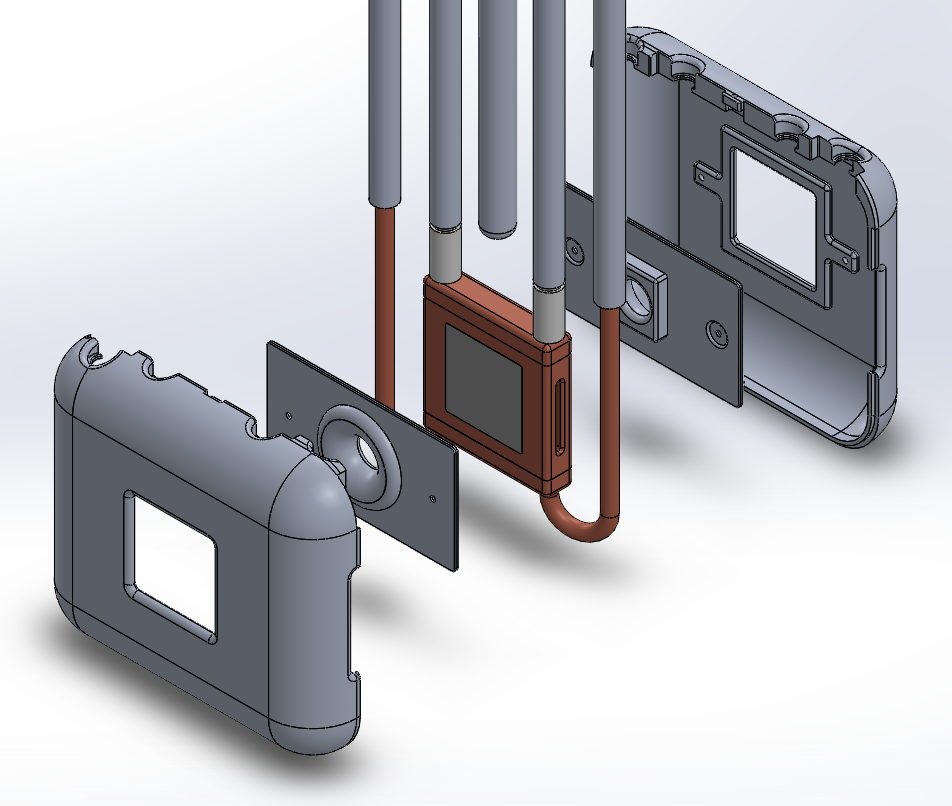}}
	\caption{Electrostatic shroud: (a) assembled, (b) exploded view}
	\label{fig:shroud}
\end{figure}

\begin{figure}
	\centering
	\includegraphics[width=0.5\textwidth]{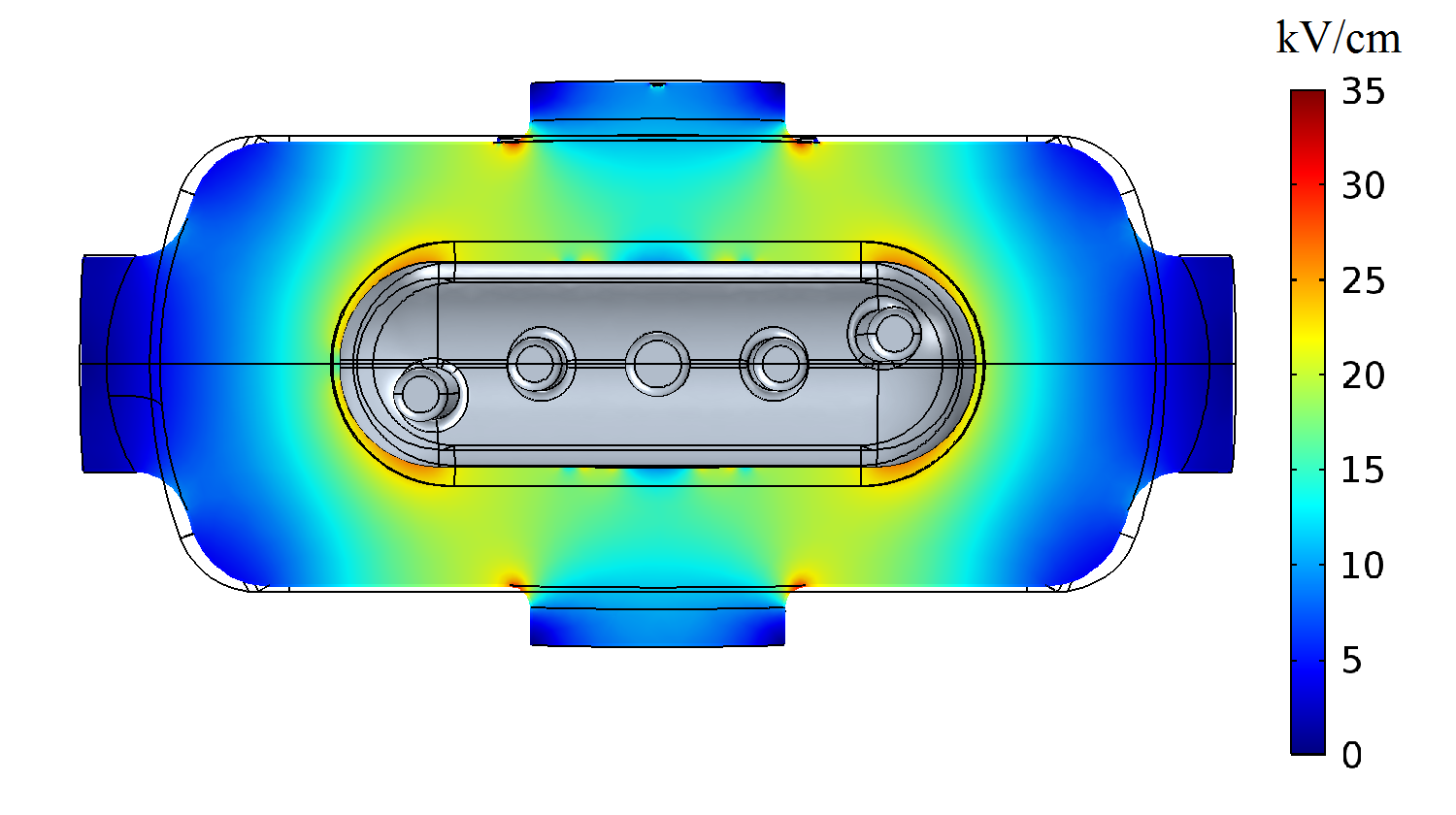}
	\caption{Electric field simulation outside shroud with target potential at -100 kV and shroud potential at -102 kV}
	\label{fig:shroud_e_field}
\end{figure}

Figure \ref{fig:shroud_e_field} shows the simulation of the electric field with the installation of the shroud design shown in Figure \ref{fig:shroud}. In the simulation, the shroud voltage is -102 kV, while the target is at -100 kV. Installation of the shroud reduced the distance between the ground and high voltage, but by designing the shroud to have a large radius of curvature everywhere, and installing insulating spacers to move the ion sources back, it was possible to limit the electric field below the 80 kV/cm electrical breakdown limit determined earlier. The maximum electric field is approximately 47 kV/cm near the top of the shroud.

\begin{figure}
	\centering
	\subfloat[\label{subfig:400V_shroud}]{\includegraphics[width=0.325\textwidth]{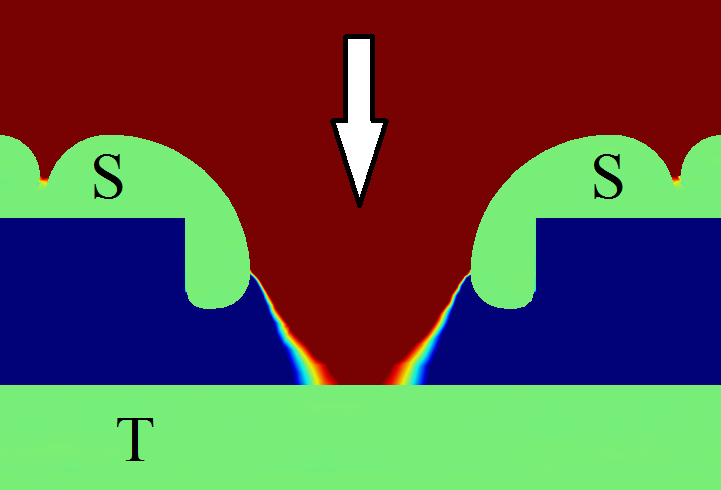}}
	\hfill
	\subfloat[\label{subfig:800V_shroud}]{\includegraphics[width=0.325\textwidth]{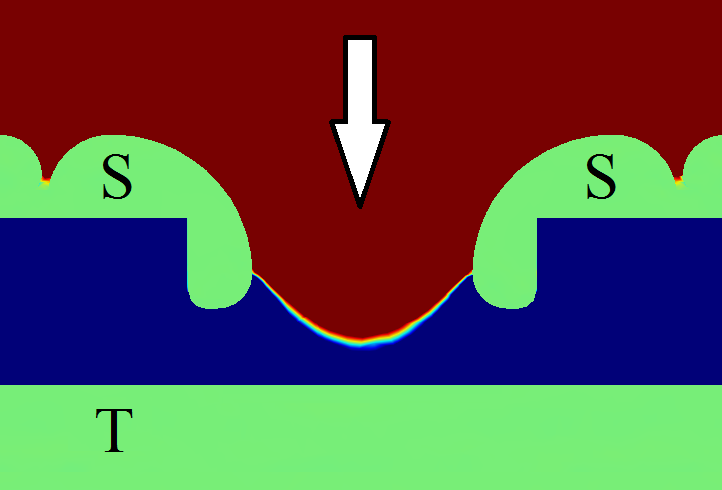}}
	\hfill
	\subfloat[\label{subfig:1400V_shroud}]{\includegraphics[width=0.325\textwidth]{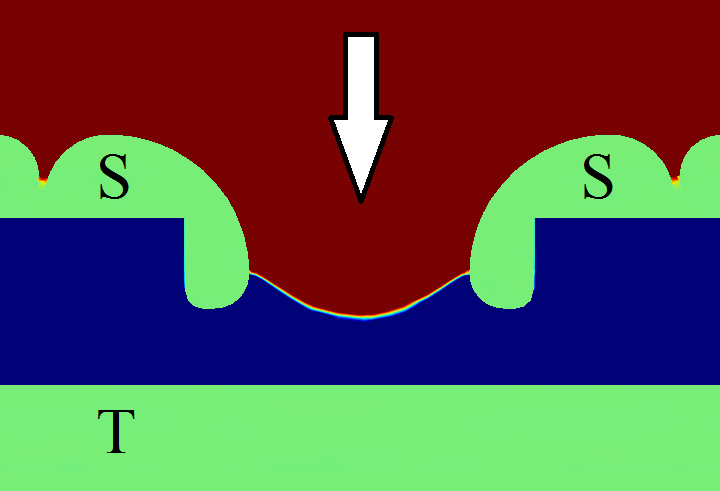}}
	\caption{Electric field direction near shroud window at (a) 400 V, (b) 800 V, (c) and 1400 V voltage differential. Red indicates field along the direction of the deuteron beam (shown by arrow), blue indicates direction opposite the deuteron beam. T and S denote the target and shroud, respectively.}
	\label{fig:shroud_e_fields}
\end{figure}

The minimum voltage differential between the target and shroud required to suppress electrons was determined by analyzing the electric field within the shroud.  Figure \ref{fig:shroud_e_fields} shows the magnitude of the electric field along the direction of the deuteron beam on the central plane within the shroud (same cross section plane shown in Figure \ref{fig:hfng_cs}). Red areas indicate regions where $E_z>0$ and blue areas indicate regions where $E_z<0$ ($z$ is the direction along the deuteron beam). For a deuteron extracted from the top of Figure \ref{fig:shroud_e_field}, the electric field will accelerate the deuteron downwards toward the target. When the deuteron enters the shroud and reaches the intersection of the red and blue region, it will have an energy of approximately 102 keV. The red region will then decelerate the deuteron to 100 keV. When the deuteron strikes the target surface and ejects electrons, the electrons will feel the electric force pushing it back toward the target.

When designing the shroud geometry, one must ensure that the red region in the top of Figure \ref{fig:shroud_e_fields} does not touch the target (as it does in Figure \ref{subfig:400V_shroud}), in order to prevent an electron from being ejected into an electric field that would accelerate the particle toward the vacuum chamber wall. Factors that affect the electric field within the shroud include the voltage differential and the distance  between the shroud and target, as well as the window size. 

\subsection{Simulation of Electron Trajectories Using Electrostatic Shroud}

\begin{figure}
	\centering
	\subfloat[\label{subfig:400V_suppression}]{\includegraphics[width=0.5\textwidth]{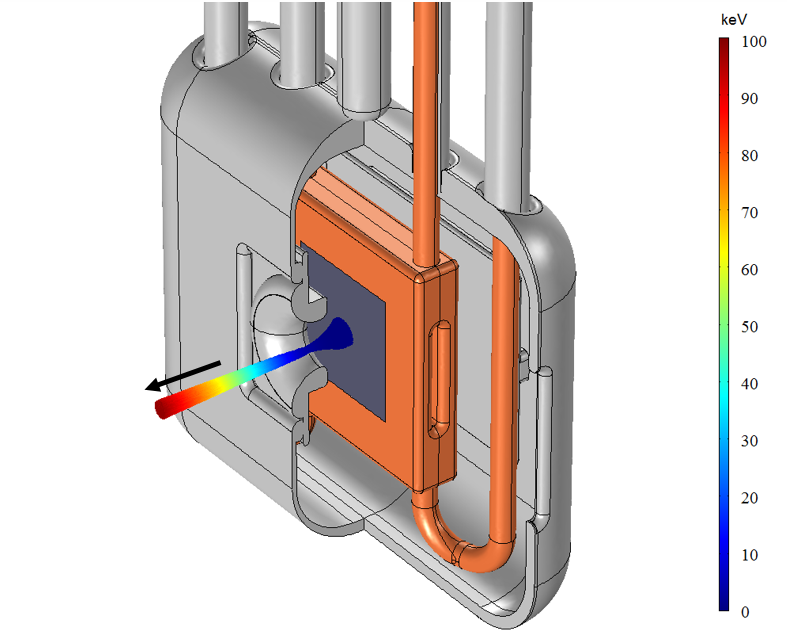}}
	\hfill
	\subfloat[\label{subfig:800V_suppression}]{\includegraphics[width=0.5\textwidth]{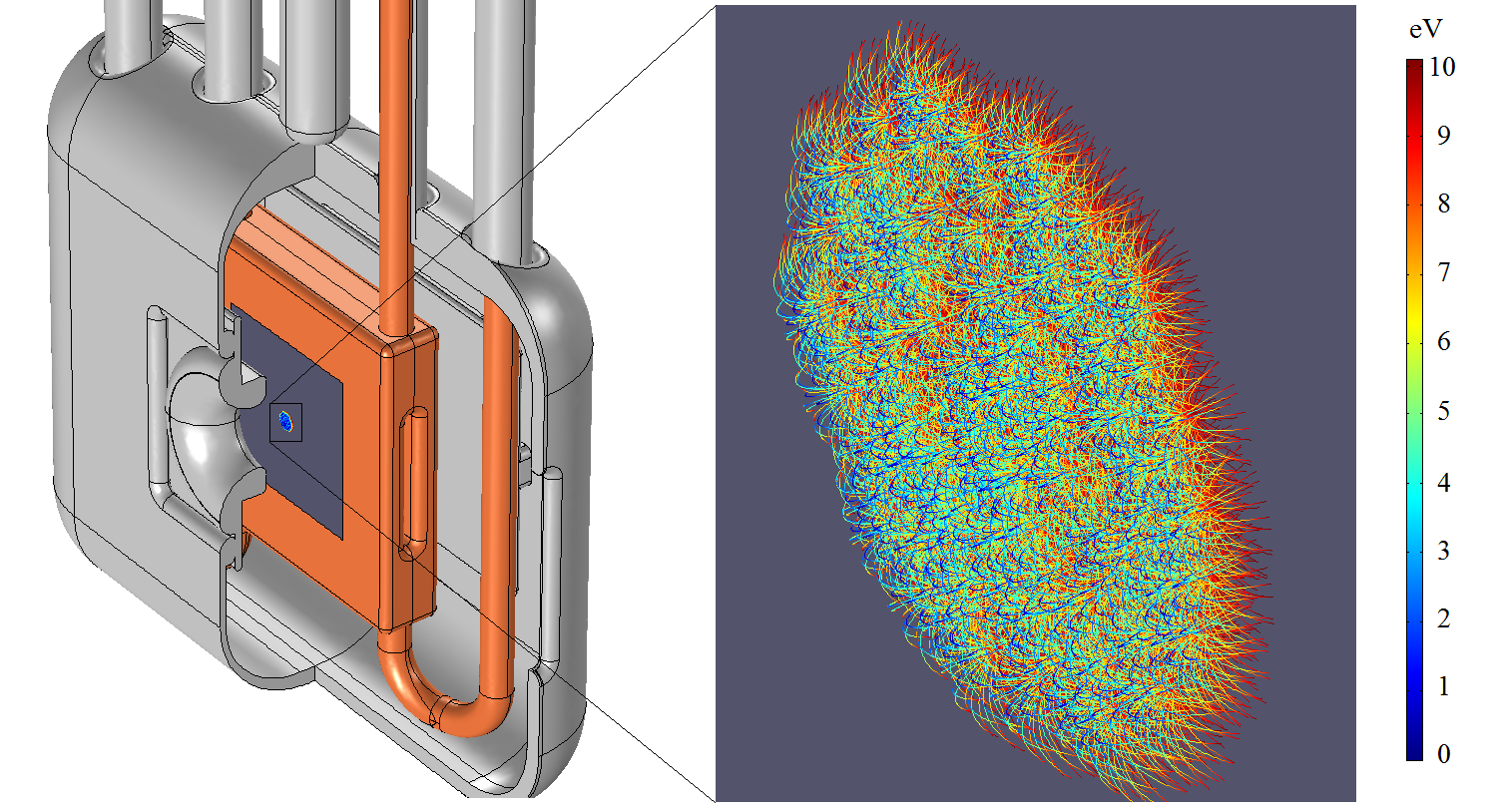}}
	\caption{Cutaway view of secondary electron trajectories with shroud differential voltages of (a) 400 V and (b) 800 V at 100 kV target potential}
	\label{fig:shroud_suppression}
	
\end{figure}

A simulation of the secondary electrons released into 2$\pi$ from a 1.3 mA deuterium beam striking the flat target surface is shown in Figure \ref{fig:shroud_suppression}. Assuming 1.2 electrons released per deuteron ion striking the target, this results in 1.56 mA of electrons. As expected and described previously due to the electric field shown in Figure \ref{subfig:400V_shroud}, for a 400 V voltage differential between the shroud and the target, secondary electrons are not suppressed and leave through the shroud window in a beam that eventually strikes the chamber with an energy of 100 keV. A shroud voltage differential of 800 V results in the electrons returning to the surface of the shroud, as seen in Figure \ref{subfig:800V_suppression}. The maximum energy the electron receives is equal to the energy at release.  

Testing of the shroud at 800 V voltage differential or greater on the HFNG has shown a large reduction in the amount of secondary electrons hitting the vacuum chamber. After implementing the shroud, no visual sign of chamber heating could be seen during machine operation due to back-streaming electrons. Also, the dose of bremsstrahlung X-rays detected by a Geiger counter in the vault decreased by a factor of more than 20 compared to the magnetic suppression technique.  

~

\section{Conclusions}

Two main methods of electron suppression were tested: using permanent magnets to bend electrons back to the target, and implementing an electrostatic shroud creating a suppressing electric field. Simulations of the permanent magnet design showed that even with magnetic fields surpassing 1000 gauss, electrons would migrate along the target due to the E$\times$B force without coming into contact with the surface. When electrons would reach an area of lower magnetic field they would escape the target area and collide with the vacuum chamber wall. Experimental tests confirmed the electrons were colliding with the vacuum chamber wall at the location shown in the COMSOL Multiphysics\textsuperscript{\textregistered} simulation.

Simulations of an electrostatic shroud showed full suppression of electrons emitted from the target surface. For the chosen shroud geometry, a voltage differential of -800 V or greater compared to the target resulted in successful electron suppression. Experimental results with the addition of a shroud reduced the back streaming electrons significantly to undetectable levels. Furthermore, the dose due to bremsstrahlung x-rays in the HFNG vault was reduced by a factor of more than 20. Analysis of both suppression techniques resulted in the conclusion that the use of electric fields for electron suppression is more effective due to the fact that electric fields add energy to the electron while a magnetic field can only change its direction.

\section{Acknowledgments}

Work supported by NSF Grant No. EAR-0960138, U.S. DOE LBNL Contract No. DE- AC02-05CH11231, and U.S. DOE LLNL Contract No. DE-AC52-07NA27344.

We would like to thank Bernhard Ludewigt, Thomas Schenkel and Qing Ji of LBNL for helpful guidance in designing the electron suppression shroud.

We also gratefully acknowledge a grant from the University of California Office of the President.
%


\bibliography{prstabbib}

\end{document}